\documentclass{aa}
\usepackage{graphicx}
\usepackage{natbib}
\bibpunct{(}{)}{;}{a}{}{,}

\begin{document}
\title{On the effect of discrete numbers of stars in chemical evolution models}

\author{M. Cervi\~no\inst{1,2} \and M. Moll\'a\inst{3}}
\institute{LAEFF (INTA), Apdo. 50727, 28080 Madrid, Spain 
          \and IAA (CSIC), Camino Bajo de Hu\'etor 24, 18080 Granada, Spain 
          \and Departamento de F\'{\i}sica Te\'{o}rica, C--XI,
                 Universidad Aut\'onoma de Madrid, 
                 28049 Cantoblanco, Spain}

\offprints{Miguel Cervi\~no}
\mail{mcs@laeff.esa.es}
\date{Received 25 February 2002; accepted 01 August 2002}

\titlerunning{Dispersion in GCE}
\authorrunning{Cervi\~no \& Moll\'a}

\abstract {We examine the impact of discrete numbers of stars in stellar
populations on the results of Chemical Evolution Models. We explore the
resulting dispersion in the true yields and their possible relation with
the dispersion in observational data based on a Simple Closed-Box model.
\\ In this framework we find that the dispersion is larger for the less
evolved or low abundance regions. Thus, the age-metallicity relation may be
a tracer of the Star Formation History of our Galaxy. This theoretical
dispersion is especially high for the relative abundance log(N/O) in
regions where the total number of stars created is still low. This may
explain part of the scatter in the N/O ratio observed in star forming
galaxies.  \\ We have also found a first order theoretical estimation for
the {\it goodness} of a linear fit of the Helium abundance vs. 12 + log
(O/H) with values of the regression coefficient between 0.9 and 0.7
(independent of sampling effects).  \\ We conclude that it is necessary to
include these sampling effects in a more realistic Chemical Evolution Model
in order that such a model reproduces, at the same time, the mean value and
the dispersion of observed abundances.  \keywords{Galaxies: abundances}}
\maketitle

\section{Introduction and motivation}

The modelling of any observable related with the integrated contribution of
stellar populations needs to assume an Initial Mass Function (IMF) and a
Star Formation History (SFH). The representation of such distributions in
the form of analytical expresions allows an easy implementation in any
evolutionary model. However, such analytical laws are valid only under the
assumption of an infinite number of stars, when sampling effects (i.e. the
integer nature of the number of stars) are not relevant. Even in the case
of the modelling of a system where the real physical properties are known,
sampling effects must be present showing a dispersion of the data around
the mean value obtained by the model \citep[see, e.g.][]{CLC00}.

\cite{CVGLMH02}, based on the work by \cite{Buzz89}, have presented a
method to evaluate quantitatively the intrinsic dispersion due to sampling
effects in spectrophotometric synthesis models. It is important to remark
that the dispersion associated to such sampling effects is indeed observed,
and used as a distance indicator, in the form of surface brightness
fluctuations in elliptical galaxies \cite[see][ for a theoretical study of
the subject]{Buzz93}.

In a next step, \cite{Ceretal01} applied this formalism to some quantities
which are time-integrated to be obtained, such as the ejected masses of
elements by the stars when they evolve.  With that work, it was
demonstrated that the estimation of these ejection rates may have a large
dispersion, mostly in the first 5 Myr of the evolution of a stellar
cluster.

On the other hand, the observational data used in the comparison of
Galactic Chemical Evolution Models (GCE) have a very large dispersion: in
the Milky Way Galaxy (MWG) there exist regions at a same Galactocentric
distance with oxygen abundances differing by 0.2 dex, means dispersions as
large as 50\%. Such dispersion can be due to observational errors, but also
to sampling effects.  As an example, a stellar cluster that has transformed
gas into stars following a Salpeter IMF in the mass range 120 -- 0.08
M$_\odot$ in a single burst, will produce $6.8 \times 10^{-3}$ SN
M$_\odot^{-1}$. In the galactic context, the most massive OB association
known \citep[Cygnus OB2, c.f.][]{Kno00} has transformed into stars $6.5
\times 10^{4}$ M$_\odot$ of gas, assuming the presence of 120 O stars in
the association (that is, in fact, an upper limit). A region of such
characteristics will produce $440\pm 20$ SN in a 63.8\% (1 $\sigma$)
confidence interval due to the limited number of stars in the region. The
formal relative dispersion is around 5\%, which will result in a similar
relative dispersion in the observed metallicity of that type of regions.
So, if the chemical evolution of our Galaxy is driven by such a type of
massive stellar clusters, at least, a {\it theoretical} relative dispersion
around 5\% must be expected in the comparison of GCE models with observed
data. But in fact, the dispersion must be larger as long as most of the
observed OB associations have a lower amount of gas transformed into stars
and the enrichment depends on the initial mass of the exploding star.

Taking into account the dispersion found for the stellar yields, we think
that the same sampling effect may also produce a dispersion in the
interstellar medium abundances. However, the stellar yields do not give
directly the abundances in a region. It is necessary to use them as the
input in a chemical evolution model.  Chemical evolution models are usually
computed through numerical methods which solve a system of equations where
the IMF such as the SFH are taken into account. However, before computing
the dispersion of the abundances predicted in these numerical chemical
evolution models, we would like to estimate the importance of these
sampling effects with simpler chemical evolution models, and see if the
resulting dispersion is at least of a similar order of magnitude to that
observed.  Therefore, the objective of this paper is to obtain a first
order estimation of how relevant these sampling effects can be in simple
GCE models in order to consider whether it is necessary to include them in
more complicated numerical chemical evolution models.

The structure of the paper is the following: In Sect. 2 we show the
application of the formalism presented in \cite{CVGLMH02} to the
computation of yields in GCE models. In Sect. 3 we obtain the dispersion on
the true yields for several metallicities and turn-off masses.  In Sect. 4
we apply the formalism to a GCE closed model and we obtain a first order
approximation of the relevance of sampling effects in the results.  We show
the conclusions and implications of this work in 5.

\section{The computation of stellar ejection rates including sampling effects}

Let us assume a system where, at a given time $t$, there are $n_{{\mathrm
d}}(m_i,t)=n_{\mathrm{d},i}$ stars of a given initial mass $m_i$ which have
completely evolved and $n_{\mathrm{a}}(m_j,t)=n_{\mathrm{a},j}$ stars of a
given initial mass $m_j$ which are still present in the system.  The total
number of completely evolved stars, $N_\mathrm{d}^{\mathrm{Tot}}(t)$, and
of stars still present in the cluster, $N_\mathrm{a}^{\mathrm{Tot}}(t)$,
are:

\begin{equation}
N_\mathrm{d}^{\mathrm{Tot}}(t) = \sum_{i=1}^{I_\mathrm{d}(t)}
n_{{\mathrm{d}},i}(t) ~~~~~;~~~~~~ N_\mathrm{a}^{\mathrm{Tot}}(t) =
\sum_{i=1}^{I_\mathrm{a}(t)} n_{{\mathrm{a}},i}(t),
\end{equation}

\noindent where $I_\mathrm{d}(t)$ and $I_\mathrm{a}(t)$ are the number of
initial masses considered and $n_{{\mathrm{d}},i}(t)$ and
$n_{{\mathrm{a}},i}$ must be integer numbers.

Let us assume that each star of mass $m_i$ is able to eject an amount of
freshly synthesized element Z given by
$my_{{\mathrm{Z}}}(m_i)=my_{{\mathrm{Z}},i}$ and defined as:

\begin{equation}
my_{{\mathrm{Z}},i}= m({\mathrm{Z}})_{\mathrm{ejected},i} - z (m_i - m_{\mathrm{rem},i})
\end{equation}

\noindent where $m({\mathrm{Z}})_{\mathrm{ejected},i}$ is the amount of the
element Z ejected, $z$ is the mass fraction of element Z in the star at the
beginning of its evolution and $m_{\mathrm{rem},i}$ is the mass of the
stellar remnant of a star with initial mass $m_i$ at the end of its
evolution.

The amount of freshly mass of the element Z ejected into the interstellar
medium at the time $t$, $Y_{{\mathrm{Z}}}(t)$ is the sum over all the
evolved stars of the individual contribution of each star:

\begin{equation}
Y_{{\mathrm{Z}}}(t) = \sum_{i=1}^{I_\mathrm{d}(t)} n_{{\mathrm{d}},i}(t)
\, my_{{\mathrm{Z}},i}
\end{equation}

The corresponding variance of $Y_{{\mathrm{Z}}}(t)$,
$\sigma^2(Y_{{\mathrm{Z}}}(t))$, is the sum of the individual
variances. Applying Poisson statistics
($\sigma^2(n_{{\mathrm{d}},i})=n_{{\mathrm{d}},i}$) we obtain:

\begin{equation}
\sigma^2(Y_{{\mathrm{Z}}}) = \sum_{i=1}^{I_\mathrm{d}}
\sigma^2(n_{{\mathrm{d}},i}) \, (my_{{\mathrm{Z}},i})^2
=\sum_{i=1}^{I_\mathrm{d}} n_{{\mathrm{d}},i} \,
(my_{{\mathrm{Z}},i})^2.
\end{equation}

The amount of material locked into stars and stellar remnants is:

\begin{equation}
M_{\mathrm{lock}}(t) = \sum_{i=1}^{I_\mathrm{d}(t)} n_{{\mathrm{d}},i}(t)\,
m_{{\mathrm{rem}},i} + \sum_{j=1}^{J_\mathrm{a}(t)} n_{{\mathrm{a}},j}(t)\,
m_j ,
\end{equation}

The corresponding variance $\sigma^2(M_{\mathrm{lock}}(t))$ is:

\begin{equation}
\sigma^2(M_{\mathrm{lock}}(t)) = \sum_{i=1}^{I_\mathrm{d}(t)}
n_{{\mathrm{d}},i}(t) \, m_{\mathrm{rem},i}^2 +
\sum_{j=1}^{J_\mathrm{a}(t)} n_{{\mathrm{a}},j}(t) \, m_j^2
\end{equation}

The yield of the element Z, $P_{\mathrm{Z}}(t)$, is defined as the fraction
of the amount of ejected material over the amount of material locked into
stars and stellar remnants:

\begin{equation}
P_{\mathrm{Z}}(t)=\frac{Y_{{\mathrm{Z}}}(t)}{M_{\mathrm{lock}}(t)}
\end{equation}

It has  an associated variance:

\begin{equation}
\sigma^2_{P_{\mathrm{Z}}}=\frac{\sigma^2(Y_{{\mathrm{Z}}})}{M_{\mathrm{lock}}^2}
+ \frac{Y_{{\mathrm{Z}}}^2 \,
\sigma^2(M_{\mathrm{lock}})}{M_{\mathrm{lock}}^4} -
2\,\frac{Y_{{\mathrm{Z}}}}{M_{\mathrm{lock}}^3} \,
{\mathrm{cov}}(M_{\mathrm{lock}},Y_{{\mathrm{Z}}}),
\end{equation}

\noindent where ${\mathrm{cov}}(M_{\mathrm{lock}},Y_{{\mathrm{Z}}})$ is the
covariance:

\begin{equation}
{\mathrm{cov}}(M_{\mathrm{lock}},Y_{{\mathrm{Z}}})=\sum_{i=1}^{I_{{\mathrm{d}}}}
n_{{\mathrm{d}},i} \, my_{{\mathrm{Z}},i} \, m_{\mathrm{rem},i}.
\end{equation}

So, at this moment we have the basic ingredients for a GCE model and their
corresponding variances.  The IMF, $\Phi(m)$ and the SFH, $\Psi(t)$, are
implicitly assumed in the computation of the $n_i(t)$ values and so, the
yield. Since our objective is obtain a first order estimation of the
relevance of the discrete number of stars (i.e. sampling effects), we will
only examine the case of the true yields.

\section{True yields}

The true yield is defined as the resulting yield of a single stellar
population.  It is in fact, the yield obtained from a generation of stars
that have suffered a single Instantaneous Burst of star formation:
$\Psi(t)=M_s \times \delta(t)$ where $M_s$ is the amount of gas that has
been transformed into stars in the IMF mass range $m_{\mathrm{up}}$ and
$m_{\mathrm{low}}$.  In this situation the values of $n_i(t)$ can be
obtained by a direct integration of the IMF:

\begin{equation}
N^{\mathrm{Tot}}_{{\mathrm{d}}}(t) = M_s \,
         \int_{m_\tau(t)}^{m_{\mathrm{up}}} \Phi(m) \, dm =
M_s \, A \, \int_{m_\tau(t)}^{m_{\mathrm{up}}} m^\alpha \, dm 
\end{equation}

\noindent where $m_{\mathrm{up}}$ is the upper limit of the IMF,
$m_\tau(t)$ is the initial mass of the star that ends its evolution at time
$t$, that can be approximated by a power law\footnote{However, see
\cite{Ceretal01} for remarks in this approximation.}: $m_\tau(t)=B\,
t^{-\gamma}$.  The IMF has been normalized to 1 M$_\odot$ with limits
$m_{\mathrm{up}}$ and $m_{\mathrm{low}}$ with a normalization constant
$A$. It has been assumed a power-law with slope $\alpha$ over all the mass
range.  Using this approximation, $N^{\mathrm{Tot}}_{{\mathrm{d}}}$ has the
functional form:

\begin{equation}
N^{\mathrm{Tot}}_{{\mathrm{d}}}(t) = \frac{M_s \, A}{\alpha + 1} 
\left( m_{\mathrm{up}}^{\alpha+1}-(B \,
t^{-\gamma})^{\alpha + 1}\right).
\end{equation}

Let us assume that $my_{{\mathrm{Z}},i}$ has a functional form
$my_{\mathrm{Z}}(m)=a m + b$.  In this case, $Y_{{\mathrm{Z}}}(t)$ and
$\sigma^2(Y_{{\mathrm{Z}}}(t))$ will have the functional form:

\begin{eqnarray}
Y_{{\mathrm{Z}}}(t) &=& M_s \int_{m_\tau(t)}^{m_{\mathrm{up}}} \Phi(m) (a
           m +b) \, dm \nonumber \\ &=& \frac{M_s \, A \, a}{\alpha+2} \left(
           m_{\mathrm{up}}^{\alpha+2}-(B \, t^{-\gamma})^{\alpha+2}\right) +
           \nonumber \\ & & \frac{M_s \, A \, b}{\alpha+1} \left(
           m_{\mathrm{up}}^{\alpha+1}-(B \, t^{-\gamma})^{\alpha+1}\right) \\
\sigma^2(Y_{{\mathrm{Z}}}(t)) &=& M_s
           \int_{m_\tau(t)}^{m_{\mathrm{up}}} \Phi(m) (a m +b)^2 \, dm
           \nonumber \\ &=& \frac{M_s \, A \, a^2}{\alpha+3} \left(
           m_{\mathrm{up}}^{\alpha+3}-(B \, t^{-\gamma})^{\alpha+3}\right) +
           \nonumber \\ & & \frac{M_s \, A \, 2ab}{\alpha+2} \left(
           m_{\mathrm{up}}^{\alpha+2}-(B \, t^{-\gamma})^{\alpha+2}\right) \nonumber
           \\ & & \frac{M_s \, A \, b^2}{\alpha+1} \left(
           m_{\mathrm{up}}^{\alpha+1}-(B \, t^{-\gamma})^{\alpha+1}\right)
\end{eqnarray}

In the same way, if we assume that $m_{\mathrm{rem}}(m)$ has also a
functional form: $m_{\mathrm{rem}}(m)=cm+d$, we may compute easily
$M_{\mathrm{lock}}$ and $\sigma^{2}(M_{\mathrm{lock}})$. Both turn out to
be functions depending on $m_{\mathrm{up}}$, $m_{\mathrm{low}}$ and $t$.

Let us now show how the relative dispersion in the different quantities
scales. At this point it is useful to define it in terms of ${\cal
N}=N_{\mathrm{eff}}$. This quantity, already used in previous works of this
series \cite[see also][]{Buzz89}, is defined, for any quantity $A$ as:

\begin{equation}
\frac{\sigma_{A}}{A}=\frac{1}{\sqrt{{\cal N}(A)}}
\end{equation}

It means that the larger $\cal N$, the lower the relative dispersion. In
our case:

\begin{eqnarray}
\frac{\sigma(Y_{{\mathrm{Z}}})}{Y_{\mathrm{Z}}} & \propto & \frac{1}{\sqrt{M_s}}
~~~{\mathrm{and}}~~~{\cal N}(Y_{{\mathrm{Z}}}) \propto M_s \nonumber \\ 
\frac{\sigma(M_{{\mathrm{lock}}})}{M_{\mathrm{lock}}} & \propto & \frac{1}{\sqrt{M_s}}
~~~{\mathrm{and}}~~~{\cal N}(M_{{\mathrm{lock}}}) \propto M_s \nonumber \\ 
\end{eqnarray}

\noindent and for the case of the true yield $p_{\mathrm Z}=
Y_{{\mathrm{Z}}} / M_{{\mathrm{lock}}}$ it can be demonstrated that
\cite[see Appendix in][ for details]{CVGLMH02}:

\begin{equation}
\frac{\sigma(p_{{\mathrm{Z}}})}{p_{\mathrm{Z}}}  \propto  \frac{1}{\sqrt{M_s}}
~~~{\mathrm{and}}~~~{\cal N}(p_{{\mathrm{Z }}}) \propto M_s 
\end{equation}

Note that $\cal N$ has the advantage of scaling always with $M_s$ whatever
the considered quantities.

With this formalism we have computed the evolution with time of the true
yields and the corresponding $\cal N$ values for Hydrogen, Helium, Carbon,
Nitrogen and Oxygen at five metallicities (Z=0.020, 0.008, 0.004 and
0.0004). We have used the ejected masses presented in \citet{Poretal98} for
massive stars (M$>$ 8 M$_\odot$). These yields include the contribution of
stellar winds and those from the type II supernova explosions. For
intermediate and low mass stars we have used the yields from
\citet{buell97}, interpolated for the above metallicities, such as those
shown in \citet{gav02}.  We have considered as solar abundances (in mass)
the values H=0.735, He=0.248, C=$2.92 \times 10^{-3}$, N=$8.56 \times
10^{-4}$ and O=$7.95 \times 10^{-3}$ taken from \citet{grev98}. The
abundances of Carbon, Nitrogen and Oxygen have been scaled linearly with
the metallicity Z for the other metallicities. The Helium abundance has
been considered as He=0.24 for metallicities different from solar.  We had
considered a Salpeter IMF slope ($\alpha=-2.35$) in the mass range 120 to
0.08 M$_\odot$.  Metallicity dependent lifetimes have been taken from
\citet{Poretal98} for all the mass range.

\begin{figure*}
\centering \includegraphics[angle=270, width=15cm]{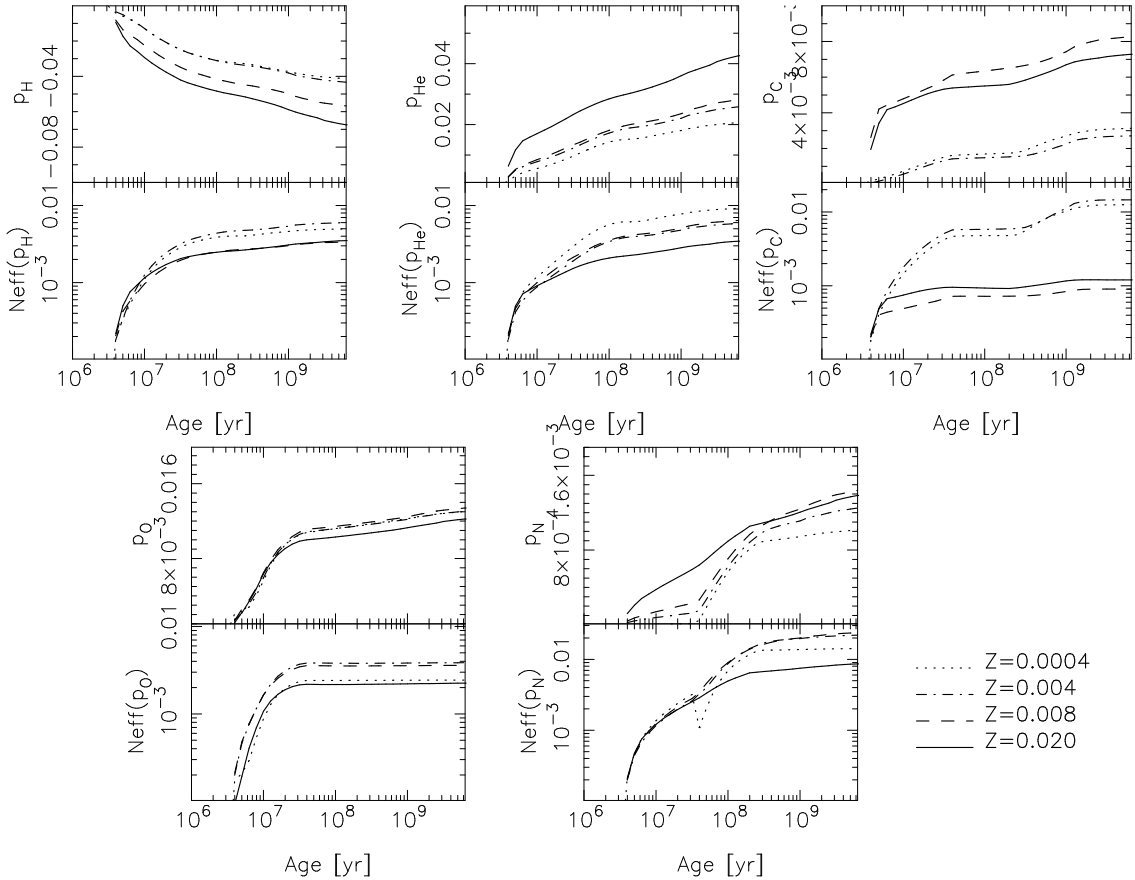}
\caption{Evolution of the true yields $p_\mathrm{H}$, $p_\mathrm{He}$,
$p_\mathrm{C}$, $p_\mathrm{O}$ and $p_\mathrm{N}$ and their relative
dispersions as a function of time.}
\label{fig:yield}
\end{figure*}

In Fig. \ref{fig:yield} we show the evolution of the true yields of
$p_\mathrm{H}$, $p_\mathrm{He}$, $p_\mathrm{C}$, $p_\mathrm{O}$ and
$p_\mathrm{N}$ and their relative dispersion as a function of time.  The
relative dispersion has been obtained including the covariance and the
dispersion in $M_{{\mathrm{lock}}}$. However, the values do not differ if
it is considered that $M_{{\mathrm{lock}}}$ has no dispersion at all,
except for ages lower than 10 Myr.  The figure shows that ${\cal
N}(p_\mathrm{O})$ reaches a constant value at $\sim$ 50 Myr, when the last
supernova explodes ($m_\tau(t=50 \mathrm{Myr})=8$ M$_\odot$).  However,
there is an evolution in the $p_\mathrm{O}$ value mainly due to the
evolution of $M_{{\mathrm{lock}}}$, that decreases with time, and a small
production of Oxygen by low mass stars (see below). The values of the true
yields $p_\mathrm{C}$ and $p_\mathrm{N}$ and their associated $\cal N$
values increase with time. This reflects the contribution of low mass stars
to the total yield, which result in an increment on the effective sources
that produce these elements and, so, a lower relative dispersion
(i.e. larger $\cal N$ values). The effect is more significant in the case
of Nitrogen, where the contribution of stars with a lifetime larger than 50
Myr (initial masses lower than 8 M$_\odot$) produces a significant
enhancement on the yield. However, the increment is similar for all the
considered metallicities, contrary to the expectations for a secondary
production of Nitrogen. On the other hand, the ejection rates used show an
strong dependence of the Carbon yield with the initial metallicity, and
hence showing a pseudo-secondary behaviour.

\begin{figure*}
\centering \includegraphics[angle=270, width=15cm]{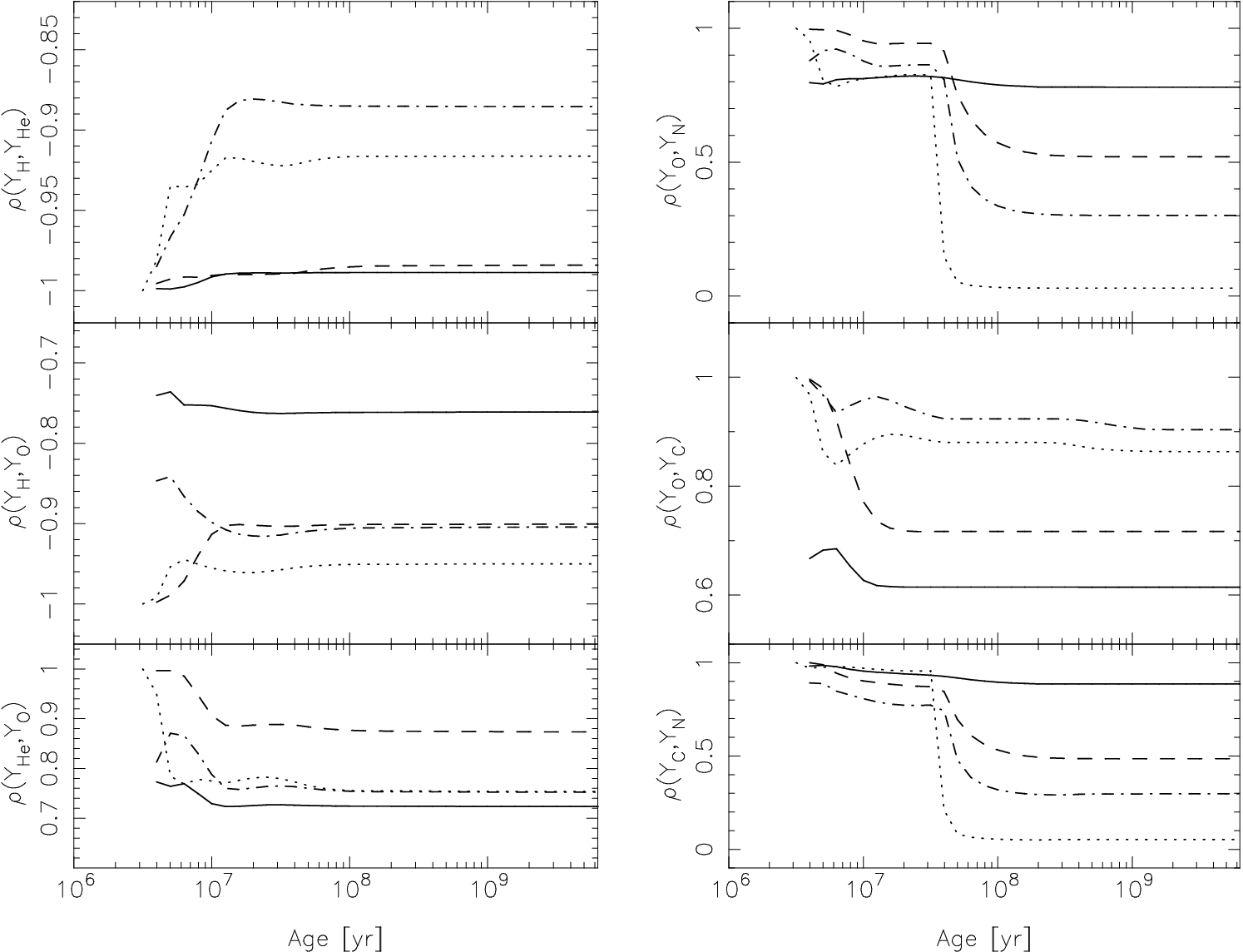}
\caption{Evolution of the correlation coefficient
$\rho(Y_\mathrm{H},Y_\mathrm{He})$, $\rho(Y_\mathrm{H},Y_\mathrm{O})$,
$\rho(Y_\mathrm{He},Y_\mathrm{O})$, 
$\rho(Y_\mathrm{O},Y_\mathrm{N})$,
$\rho(Y_\mathrm{O},Y_\mathrm{C})$, and $\rho(Y_\mathrm{C},Y_\mathrm{N})$
as a function of time. Symbols like in Fig. \ref{fig:yield}.}
\label{fig:cov}
\end{figure*}

We show the evolution of the correlation coefficient of
$\rho(Y_\mathrm{H},Y_\mathrm{He})$, $\rho(Y_\mathrm{H},Y_\mathrm{O})$,
$\rho(Y_\mathrm{He},Y_\mathrm{O})$, $\rho(Y_\mathrm{O},Y_\mathrm{N})$,
$\rho(Y_\mathrm{O},Y_\mathrm{C})$, and $\rho(Y_\mathrm{C},Y_\mathrm{N})$ in
Fig. \ref{fig:cov}. Note that the value of the covariance coefficient is an
absolute one, independent of sampling effects. In this case it is a measure
of the productions of the elements among different types of stars.

The figure shows that $Y_\mathrm{H}$ and $Y_\mathrm{O}$, and $Y_\mathrm{H}$
and $Y_\mathrm{He}$ are almost anti-correlated for all the ages and
metallicities. In a first order approximation, it means that the regression
coefficient (i.e. the correlation coefficient) of a linear fit of the
abundance of Helium or Oxygen vs. abundance of Hydrogen will be close to
-1. In the case of $\rho(Y_\mathrm{He},Y_\mathrm{O})$ the figure shows that
a linear fit of the Helium abundance vs. 12 + log (O/H), as the one used
for the determination of the primordial Helium abundance, may have a value
between 0.9 and 0.7 (i.e. there will be an intrinsic dispersion in such
type of representation of the data, as it is indeed observed).  However, we
want point out that we are working with true yields, which are different
from the observed ones.

In the case of $Y_\mathrm{O}$ and $Y_\mathrm{N}$ the correlation is
relatively strong (close to 1) in the first 50 Myr, when the production of
Oxygen and Nitrogen is due to a similar population of stars. For ages
larger than 50 Myr, the correlation coefficient becomes dependent on the
metallicity: the lower the metallicity, the closer to zero the correlation
coefficient will be. A zero value in the correlation coefficient means that
Oxygen and Nitrogen are produced by different types of stars. It also
means, in a first order approximation (but see next section), that the N/O
ratio might have a larger dispersion in the low abundance regions, when the
contamination of the ISM is due to stars with an initial low metallicity.

The correlation between Carbon and Oxygen shows a different behavior. The
correlation coefficient is also dependent on the initial metallicity, but
in this case the lower value is 0.6, i.e. the production of Oxygen and
Carbon is produced by similar (but not the same) stellar populations.
Also, the lower value of the correlation coefficient at solar metallicity
implies that, in a first order approximation, the dispersion in the C/O
ratio will increase with metallicity.

Finally, we had also plotted the correlation between Carbon and Nitrogen,
very similar to the one for Oxygen and Nitrogen. This is the expected
behaviour if Carbon and Oxygen originate mostly from massive stars.  Such
a correlation coefficient was also shown in \citet{Ceretal01} for solar
metallicity and ages between 1 and 10 Myr both being quite similar in both
cases despite the different treatment used.

Once these true yields, their time evolution when a simple stellar
population is created, and their dispersions have been computed, we can use
them for calculating the abundances of these elements. As we have already
explained, we want to use simple models before treating this subject with
numerical models, thus we will use these yields in the most simple GCE
model: The Closed-Box Model.

\section{The Closed-Box Model}

The original formalism of the Closed-Box model was established by
\citet{tin80}. \citet{Mae92} revised the formalism and we follow the
prescriptions of this author.  Let us assume that the total mass in the
system, $M$, is constant: $M=s+g=\mathrm{c}$, where $g$ is the amount of
mass in the gas and $s$ the amount of mass in stars.  In this case

\begin{equation}
Z(t) - Z_0= p_{\mathrm{Z}} \ln \frac{g+s}{g}=p_{\mathrm{Z}} \ln \mu^{-1}=-
p_{\mathrm{Z}}(t) \ln \mu(t),
\end{equation}

\noindent where $\mu$ is the gas fraction with an associated variance
$\sigma^2_\mu=\sigma^2_s/M^2$, and $Z_0=Z(t=0)$ is the metallicity at the
onset of the burst of star formation.

\begin{figure*}
\centering \includegraphics[angle=270, width=13cm]{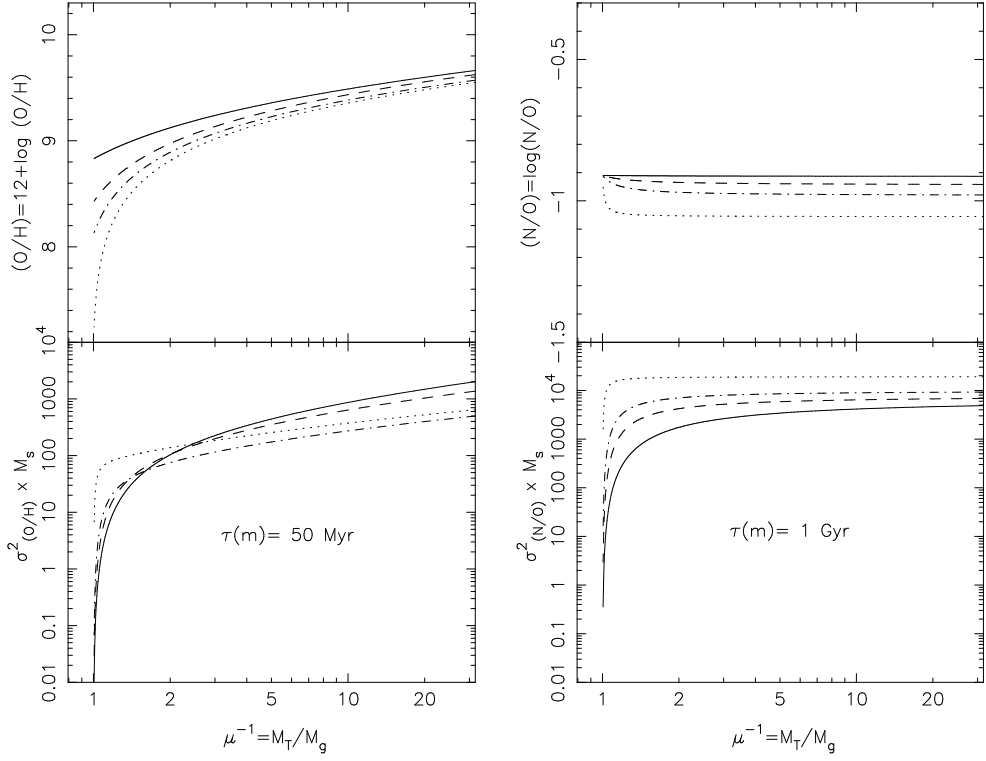}
\caption{Evolution of the $12+log(O/H)$ and $log(N/O)$ ratios as a 
 function of $\mu$, with $\tau(m)$=50 Myr, for oxygen and $\tau(m)$=1 Gyr, 
for nitrogen. Symbols like in Fig. \ref{fig:yield}.}
\label{fig:Z}
\end{figure*}

For simplicity, let us assume that there is only a dispersion due to the
discrete number of stars in $Y_{\mathrm{Z}}$. In this case, the correlation
coefficient $\rho(Y_{\mathrm{Z_1}},Y_{\mathrm{Z_2}})$ equals to
$\rho(p_{\mathrm{Z_1}},p_{\mathrm{Z_2}})$.

The variance in $Z$ is then:

\begin{equation}
\sigma^2_Z=\left(- \ln \mu\right)^2
   \sigma^2(p_{{\mathrm{Z}}})
\end{equation}

And the relative dispersion of the abundance ratio of two elements $Z_1$
respect to $Z_2$ is:

\begin{eqnarray}
\frac{\sigma^2_{Z_1/Z_2}}{(Z_1/Z_2)^2}&=&\ln^2\mu\left[\frac{\sigma^2_{p_{\mathrm{Z_1}}}}{Z_1^2}+
\frac{\sigma^2_{p_{\mathrm{Z_2}}}}{Z_2^2}-\frac{2\,
  \mathrm{cov}(p_{{\mathrm{Z_1}}},p_{{\mathrm{Z_2}}})}{Z_1\, Z_2}\right] 
\nonumber \\
\label{eq:disp2}
\end{eqnarray}

Note that the dispersion scales with $\ln^2\mu$. In a first order
approximation it means that the dispersion on ratios of elements will be
larger for low metallicity systems.

The Closed-Box Model needs the use of an additional assumption about the
true yield. In such model, the time evolution of the true yield is not
considered, because, due to the hypotheses involved in its calculation, the
true yield corresponds to the total yield obtained for a single stellar
generation, that is, the maximum value obtained when all stars die. This is
the final value reached at the end of the time evolution in
Fig.~\ref{fig:yield}. Thus, the abundance computed with the Closed-Box
Model is an upper limit of this one which may be actually observed in a
star forming region. On other hand the Closed-Box model assumes that all
stars are divided in two groups, a first one formed by the low mass stars
living forever, and a second one with the more massive which will die
immediately after being created. The mass limit between both groups depends
on time. Therefore, to use adequately the Closed-Box model, it is necessary
define a time scale (or a turn-off mass) $\tau(m)$ where it is assumed that
all the stars with lifetimes shorter than $\tau(m)$ (or masses larger than
$m$) produce all their yield instantaneously.  Concerning the total mass
transformed into stars, the Closed-Box model is not dependent on the
assumed SFH, so, an amount of mass $M_s$ transformed into stars or a
constant star formation rate with a value $M_s/\tau(m)$ produce the same
result.  Finally, it is also needed a value of $\mu$ to describe the
evolution of the abundance. In the following we will use a value of
$\tau(m)= 50$ Myr to obtain the dispersion in the age--O/H relation and an
value of $\tau(m)= 1$ Gyr to obtain the dispersion on the N/O ratio. The
metallicity evolution will be performed in terms of $\mu$.  Finally, we
have assumed the same $Z_0$ values than the ones used for the computation
of the true yields.

\subsection{Age--O/H relation}

Let us now consider the age-metallicity relation. Although this relation is
usually given as a function of [Fe/H], we will use the abundance (O/H) as
the metallicity indicator. To calculate the Iron yield it is necessary to
include the ejection from type I SN explosions, which are not considered in
this work.

The stellar population has produced most of the Oxygen at 50 Myr, so we
have chosen this value of $\tau(m)$ to obtain the dispersion in the
age--O/H relation. We have transformed the mass-abundance ratio in the
element-abundance ratio by number obtaining this in the used way:

\begin{eqnarray}
(O/H) = 12 + \log \frac{O}{H} = 12 + \log \frac{Z_O}{16 \times Z_H}
\end{eqnarray}

The corresponding dispersion as a function of $\mu$ is shown in the left
panel Fig. \ref{fig:Z}.  The dispersion increases with lower values of
$\mu$ as is expected from Eq. (\ref{eq:disp2}).

As an example, let us assume the yield at Z=0.0004. It reaches a value of
$12+\log(O/H)=8.93$ at $\mu=0.4$. Let us also assume a constant star
formation rate of $\Psi(t)=4$ M$_\odot$ pc$^{-2}$ Gyr$^{-1}$ for the solar
cylinder \cite[c.f.][ Table 7.7]{Pagel}.  In this situation, a dispersion
of $\sigma_{(O/H)}=28.5$ pc$^{-1}$ dex is expected, i.e. $\sim$ 0.3 dex if
the star formation takes place in regions of 100$\times$100 pc$^2$.

This value is similar to the observational errors.  In general, the
dispersion will scale with the amount of gas transformed into stars or with
the star formation rate (once defined a time scale).  So, in this simple
framework, the scatter in the age--metallicity relation is related with the
Star Formation History itself. Unfortunately, this statement can be only
verified with more sophisticated GCE, but that is not the goal of this
paper.

It is also interesting to note that the recompilation of correlation
coefficients of the fit of $O/H$ versus Galactocentric distance for the MWG
presented in \citet{hen99} has a composite value of -0.63. In the case of
the true yields, the correlation coefficient between $Y_\mathrm{H}$ and
$Y_\mathrm{O}$ varies between 0.75 (Z=Z$_\odot$) and 0.95 (Z=0.004) with a
strong dependence on metallicity. In our framework is not possible to
compute the correlation coefficient of the relation between $O/H$ and the
Galactocentric distance, but a more elaborated model must be able to obtain
a theoretical value of this correlation coefficient comparable with the
observed one.

\subsection{N/O ratio}

For the N/O ratio, we had chosen a value of $\tau(m)=1$ Gyr that implies
that all the stars with mass larger than $\sim$ 2 M$_\odot$ have enriched
the ISM with their Nitrogen. The N/O ratio has been computed with the
relation:

\begin{eqnarray}
(N/O) &=& \log \frac{N}{O} = \log \frac{14 \times Z_O}{16 \times Z_N}
\end{eqnarray}

The corresponding dispersion as a function of $\mu$ is shown in the right
panel of Fig. \ref{fig:Z}. Again, a direct comparison with observed data
can not be performed. First of all, the $(N/O)$ value is fixed by the
yields used (and the $\tau(m)$ value). However, it can be shown that a
$\sigma_{(N/O)}$ value larger than 0.5 dex will be a natural effect if the
ISM has been contaminated by a burst of star formation with an amount of
gas transformed into stars lower than $8 \times 10^4$ M$_\odot$ or,
equivalently, star formation rates lower than $\Psi(t)=80 $ M$_\odot$
Myr$^{-1}$ (using the yield for Z=0.0004). Again, a more sophisticated
model is needed to establish any conclusion.
 
\section{Conclusions}

In this work we have shown a theoretical formalism for the evaluation of
the dispersion on GCE models. Using a simple Closed-Box model, we have
obtained a first order approximation to the relevance of the discreteness
of the stellar population and the sampling effects in GCE models.  Despite
the approximations, the effect of the discreteness of the stellar
populations may be relevant for explaining the dispersion in the
age--metallicity relation. Such observed dispersion may be an indicator of
the SFH of our Galaxy. It may be also relevant in the study of the observed
dispersion in the N/O ratio in H{\sc ii} regions and dwarf galaxies.

We have found that the dispersion in the N/O ratio is larger for lower
metallicities. This effect is indeed observed in our Galaxy.  We have
obtained a first order theoretical estimation for the {\it goodness} of a
linear fit of the Helium abundance vs. 12 + log (O/H) with values of the
regression coefficient between 0.9 and 0.7.

Finally, we want to note that in the case of GCE models attempting to trace
the metallicity distribution of our Galaxy, the dispersion will not scale
with the total mass of the Galaxy. Instead, it must be computed with the
masses of the subsystems that are sampled: the dispersion of abundances for
individual H{\sc ii} regions from the global metallicity gradient observed
in our Galaxy \citep{hen99} could be understood in this framework since
these regions are less massive systems.

However, no conclusion can be obtained unless this effect is included in a
more sophisticated GCE.  It is therefore necessary to make a more
exhaustive study on this subject. It will be performed in forthcoming
papers.

\acknowledgements We want to acknowledge Prof. Chiosi and V.  Luridiana for
their useful suggestions which motivated us to write this paper. We also
want to acknowledge V. Luridiana, J.M. V{\'{\i}}lchez and L. Carigi for
useful discussions.

\bibliographystyle{apj}

\end{document}